# Ultralow Lattice Thermal Conductivity in Orthorhombic $Rb_2Se_3$ with rigid covalent trimers and a soft ionic framework

Tiantian Jia[1,*], Jinxiang Liu[1], Yongsen Tang[1], Shengli Liu[1,*] and Yongsheng Zhang[2,*]
[1]School of Science, Nanjing University of Posts and Telecommunications (NUPT), Nanjing 210023, China
[2]Advanced Research Institute of Multidisciplinary Sciences, Qufu Normal University, Qufu 273165, China

Corresponding author:
ttjia@njupt.edu.cn (Tiantian Jia)
liusl@njupt.edu.cn (Shengli Liu)
yshzhang@qfnu.edu.cn (Yongsheng Zhang)

Alkali-metal-containing compounds often exhibit exceptionally low lattice thermal conductivity, yet the structural origins and transport mechanisms remain incompletely understood. Here, we investigate orthorhombic $Rb_2Se_3$ using first-principles bonding, lattice-dynamical, and thermal-transport calculations. Bonding analyses reveal strongly covalent, V-shaped $Se_1$–$Se_2$–$Se_1$ trimers embedded in a soft ionic coordination framework. Representing each trimer as a pseudo-atom reveals a pseudo-one-dimensional chain arrangement along [001] within the three-dimensional crystal. This bonding hierarchy produces a dense low-frequency phonon manifold and four separated high-frequency optical branches dominated by internal trimer stretching. Four-phonon scattering strongly suppresses particle-like heat transport, leaving the coherent contribution dominant. The coherent contribution is largest along the pseudo-atom-chain direction [001] and accounts for most of the thermal conductivity anisotropy. At 300 K, the total lattice thermal conductivities are 0.067, 0.096, and 0.265 W/mK along [100], [010], and [001], respectively. These results connect the coexistence of rigid covalent trimers and a soft ionic framework to ultralow thermal conductivity, while revealing how their structural arrangement is reflected in directional coherent heat transfer.

## INTRODUCTION

Low lattice thermal conductivity ( $\kappa_l$ ) is important for high-performance thermoelectric energy conversion and high-temperature thermal barrier coatings. According to the Boltzmann transport theory [1], the lattice thermal conductivity in semiconductors is primarily determined by their phonon properties. Low lattice thermal conductivity can arise from small phonon group velocities or short lifetimes. Numerous studies have shown that materials with distinctive crystal structures and atomic bonding configurations tend to possess low $\kappa_l$. Examples include compounds with cage-like structures, such as filled skutterudites [2,3], and clathrates [4,5]; layered compounds such as SnSe [6,7] and $CuP_2$ [8]; lone-pair-containing materials like InTe [9], $TlInTe_2$ [10], and $NaSbSe_2$ [11]; resonantly bonded materials such as PbTe [12] and phosphorene [13]; and compounds containing rattling atoms, exemplified by Cu-based [14,15] and Ag-based [16,17] compounds. Phonon analyses of these systems link their low $\kappa_l$ to soft vibrations and anharmonic interactions associated with particular atomic or structural units.

Using a high-throughput evaluation method, we [18] previously investigated the thermoelectric properties of 243 known binary semiconductor chalcogenides and identified candidates combining promising electrical transport properties with ultralow $\kappa_l$. Recently, using a simple structural rattling descriptor, Li et al. [19] identified 532 materials with rattling atoms exhibiting ultralow $\kappa_l$. Analysis of these predicted materials revealed a common compositional feature: the majority contain alkali metals. $Rb_2Se_3$ was identified in both screenings and has a relatively simple orthorhombic structure, making it a useful system for examining the origin of low $\kappa_l$ in alkali-metal-based compounds. Experimentally, $Rb_2Se_3$ was synthesized and structurally characterized by P. Böttcher in 1980 [20]. Theoretically, the high-throughput first-principles phonon calculations of Petretto et al. [21] showed that the phonon dispersions of $Rb_2Se_3$ exhibited no imaginary frequencies, supporting harmonic dynamical stability.

Alkali metals (such as Li, Na, K, Rb and Cs) are characterized by their low ionization energies, enabling them to readily lose their valence electrons and form cations. This predisposition favors the formation of ionic bonds with nonmetals, resulting in compounds with predominantly ionic character. Recent studies have identified low-energy vibrational modes and antibonding states in the valence bands of several alkali-metal-containing compounds. These features have been associated with low lattice thermal conductivity in systems such as XS (X = K, Rb, and Cs) [22], $CsPbBr_3$ [22], $CsCu_3S_2$ [23], KYBi (Y=Sr, Ca) [24,25], etc. [26–28] These findings suggest a correlation between soft phonon properties, strong anharmonicity, and low $\kappa_l$ in alkali-metal-based materials. Understanding how the bonding environment produces these vibrations is therefore important for interpreting their thermal transport.

A structural explanation of low conductivity requires understanding how the associated vibrations carry and scatter heat. Two aspects of the transport calculation are relevant here. First, conventional three-phonon calculations account for anharmonic scattering through cubic interatomic force constants [22,29,30]. In strongly anharmonic crystals, four-phonon (4ph) processes arising from quartic force constants can provide substantial additional broadening, with rates comparable to or larger than those of three-phonon (3ph) processes [26,28]. Second, when phonon lifetimes are short and neighboring branches are closely spaced, the Peierls–Boltzmann description of independent phonon carriers can become insufficient. The Wigner formulation includes both a particle-like contribution associated with diagonal heat-current terms and a coherent, wave-like contribution associated with off-diagonal coupling between phonon branches [27,31]. Together, these considerations show that higher-order scattering and interbranch coherence can substantially affect heat transport when anharmonic broadening is strong and phonon branches are closely spaced. The exceptionally low lattice thermal conductivity predicted for $Rb_2Se_3$ motivates testing whether a three-phonon, particle-like description is sufficient. We therefore evaluate four-phonon scattering and coherent heat transfer to establish how they affect the conductivity.

In this work, we investigate how the bonding hierarchy and structural organization of $Rb_2Se_3$ govern its particle-like and coherent heat transport. Strongly covalent, V-shaped $Se_1$–$Se_2$–$Se_1$ trimers are embedded in a softer ionic coordination framework. Representing each trimer as a pseudo-atom reveals a pseudo-one-dimensional chain arrangement along [001] within the three-dimensional crystal. We examine how this bonding hierarchy separates internal trimer vibrations from collective lattice motion, and whether the chain direction is reflected in the particle-like and coherent heat currents. The calculations show that four-phonon scattering strongly suppresses particle-like conduction along all three axes, whereas coherent transfer dominates the remaining heat flow and is largest along [001]. This connects the coexistence of stiff covalent units and soft collective motion to ultralow conductivity, while identifying the coherent contribution as the main source of its anisotropy. This work provides atomic-level insights into anharmonicity in polyanion alkali selenides and delivers a transferable design rule for intrinsic glasslike thermoelectric ionic materials.

## COMPUTATIONAL METHODOLOGIES

First-principles calculations were performed based on density-functional theory (DFT) with the projector augmented-wave (PAW) method [32], as implemented in the Vienna Ab initio Simulation Package (VASP) [33]. To model exchange-correlation interactions, the semilocal generalized gradient approximation (PBEsol) [34] functional was employed because it provides a suitable description of equilibrium structures and phonon frequencies in solids [35]. Full structural optimizations, including both atomic positions and lattice parameters, were carried out using the conjugate gradient method. The production calculations used the standard PAW datasets for Rb ($4s^2 4p^6 5s^1$) and Se ($4s^2 4p^4$) supplied with the VASP distribution. The convergence criteria were set to $10^{-8}$ eV for total energies and $10^{-4}$ eV/Å for residual atomic forces. A plane-wave cutoff energy of 600 eV was selected, and the Brillouin zone was sampled using a 7 × 7 × 7 Monkhorst-Pack k-point grid [36]. Bader partitioning was used to quantify charge redistribution [37–40]. Energy-resolved crystal orbital Hamilton population (COHP) curves and their cumulative integrals (ICOHP) were calculated using LOBSTER [41,42]. The values integrated up to the Fermi level, ICOHP($E_f$), were used to compare the net bonding contributions of different atomic pairs. Larger positive −ICOHP($E_f$) values indicate greater net bonding stabilization within the orbital projection. The electron localization functions (ELFs) [43] were used to examine the spatial distribution of localized electrons.

For the harmonic calculations, the phonon vibrational properties of $Rb_2Se_3$ were computed from second-order force constants obtained by density functional perturbation theory (DFPT) in VASP. The phonon calculations included the nonanalytic dipole correction based on Born effective charges and dielectric tensors [44,45]. PHONOPY [46] was used to construct the phonon dispersion. The structure was reoptimized using a denser $9 \times 9 \times 9$ Monkhorst-Pack k-point grid. The second-order interatomic force constants (IFCs) were calculated using $3 \times 3 \times 3$ supercells containing 270 atoms. To validate the reliability of the calculated phonon vibrational properties, we adopt the meta-GGA $r^2$SCAN [47] [48], exchange-correlation functional for cross-checking our PBEsol results. The mode-resolved quasi-harmonic Grüneisen parameters were obtained from phonon calculations at the equilibrium volume and at volumes scaled by ±3%, using the finite-difference relation $\gamma(q) = -\partial \ln \omega(q) / \partial \ln V$ in PHONOPY. For comparison, mode-resolved Grüneisen parameters evaluated explicitly from the third-order interatomic force constants were computed within the ShengBTE framework [49]. The finite-volume and third-order-IFC calculations used $31 \times 31 \times 31$ and $7 \times 7 \times 7$ q-point meshes, respectively. The finite-volume structures include changes in cell shape and internal coordinates.

With the cutoff distance for interatomic interactions set to include up to the fifth and second nearest-neighbor shells for third-order and fourth-order IFCs, respectively, the third-order and fourth-order IFCs were determined using the finite displacements method via the thirdorder.py [49] and fourthorder.py [50] codes. The lattice thermal conductivity was decomposed within the Wigner transport formalism [51,52]: $\kappa_l = \kappa_p + \kappa_c$, where $\kappa_p$ is the particle-like Peierls-Boltzmann thermal conductivity (diagonal heat flux terms, phonon quasiparticle transport), and $\kappa_c$ is the wave-like coherent thermal conductivity associated with off-diagonal coupling between phonon branches. Calculations were performed using ShengBTE [49] coupled with the FourPhonon extension module [53] and the Wigner transport module [31,51,52,54,55] to incorporate both three-phonon and four-phonon anharmonic scattering.

A Γ-centered $7 \times 7 \times 7$ q-point mesh and a Gaussian broadening scale factor of 1.0 were used. Both the four-phonon phase-space and scattering-rate sampling parameters were set to $5 \times 10^4$. The three-phonon collision operator was treated iteratively, with four-phonon scattering included as a diagonal relaxation term. Natural-abundance isotope scattering was also included. For each scattering model, the particlelike and coherent contributions were evaluated using consistent phonon properties and scattering inputs. The linewidths entering the coherent term included three-phonon and isotope scattering, with four-phonon scattering added in the corresponding model. The calculations describe bulk crystals and omit boundary and extrinsic defect scattering. The relative convergence tolerance for the iterative solution was $10^{-8}$. The q-point mesh sensitivity tests and the three-phonon comparison with almaBTE [56] are presented in Figs. S1 and S2 of the Supplemental Material, respectively. Table SI compares the particle-like RTA and iterative results at 300 K.

## RESULTS AND DISCUSSION

### A. Crystal structure and bonding hierarchy

The crystal structure of $Rb_2Se_3$ adopts the orthorhombic space group $Cmc2_1$, as illustrated in Figs. 1(a) and 1(b), depicting the conventional unit cell and primitive cell structures, respectively. The calculated lattice constants (a = 7.595 Å, b = 10.994 Å, c = 7.722 Å) are in reasonable agreement with the experimental data (a = 7.856 Å, b = 10.858 Å, c = 7.977 Å) [20]. Two inequivalent Rb sites ($Rb_1$, $Rb_2$) and two inequivalent Se sites ($Se_1$, $Se_2$) occupy the unit cell. The coordination numbers are 7, 8, 7, and 5, for $Rb_1$, $Rb_2$, $Se_1$ and $Se_2$, respectively. The bonding configurations of these atoms are detailed in Fig. 1(c). Each $Rb_1$ atom is coordinated by six $Se_1$ atoms and one $Se_2$ atom, while each $Rb_2$ atom is coordinated by six $Se_1$ atoms and two $Se_2$ atoms. Each $Se_1$ atom bonds to three $Rb_1$ atoms, three $Rb_2$ atoms, and one $Se_2$ atom. Each $Se_2$ atom bonds with one $Rb_1$ atom, two $Rb_2$ atoms, and two equivalent $Se_1$ atoms.

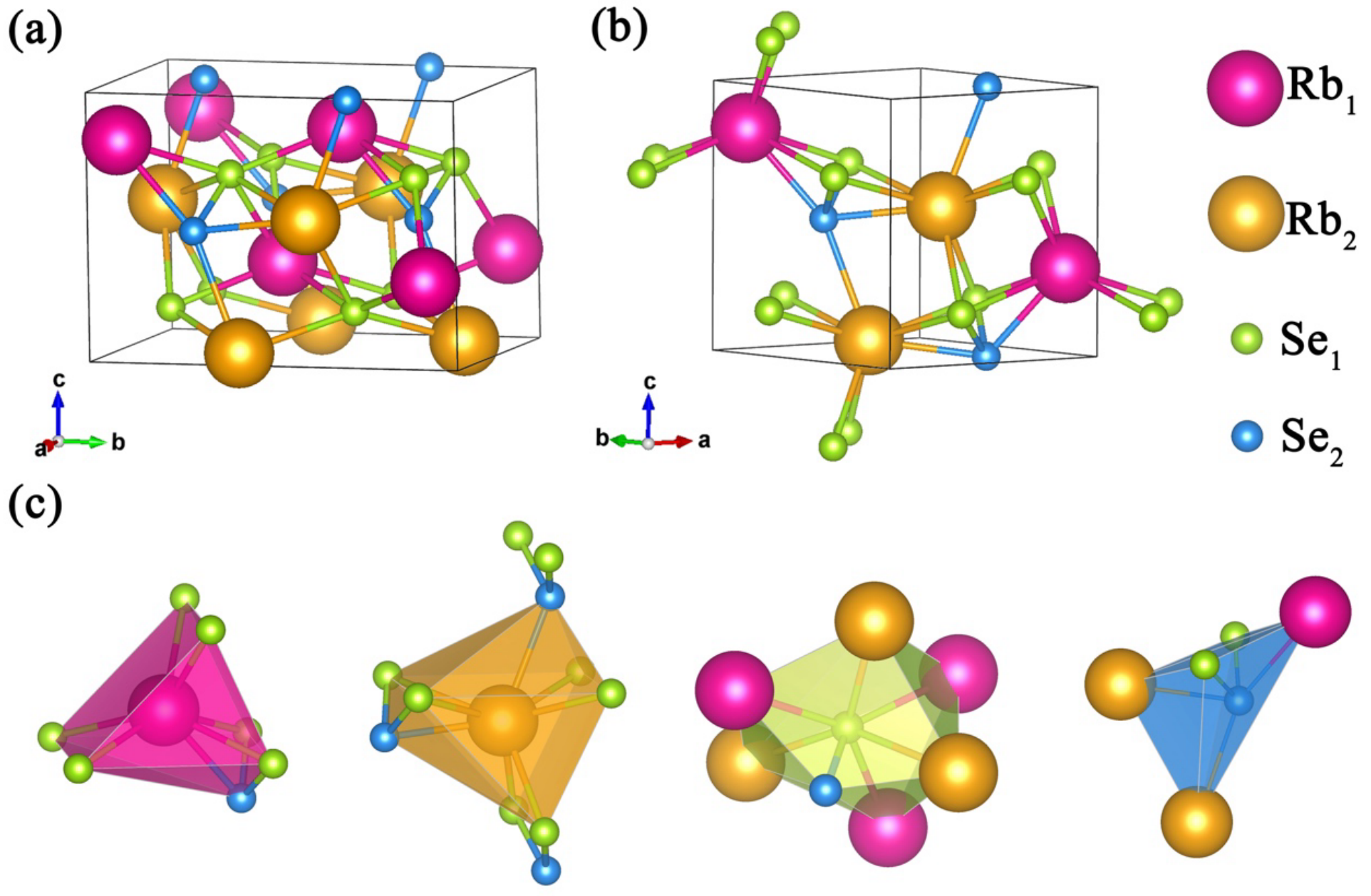

Figure 1. The conventional unit cell structure (a), primitive cell structure (b), and the bonding configurations of the different atoms (c) in $Rb_2Se_3$.

To quantitatively characterize the bonding, we performed Bader charge and COHP analyses. The Bader charge transfer ($\Delta q$) was calculated by subtracting the Bader electron population ($q_B$) from the valence-electron count specified by the corresponding PAW dataset ($q_Z$), as $\Delta q = q_Z - q_B$. Here, positive $\Delta q$ denotes electron donation and negative $\Delta q$ denotes electron accumulation. The calculated $\Delta q$ for the different atoms, together with the bond lengths ($L_{ij}$) and the −ICOHP($E_f$) values are summarized in Table I. The calculated energy-resolved −COHP and cumulative −ICOHP curves are presented in Fig. S3 of the Supplemental Material.

Several important features emerge from these data. The $Se_1$–$Se_2$ distance is 2.411 Å, and its −ICOHP($E_f$) value of 3.727 is more than ten times those of the Rb–Se contacts. By comparison, the $Se_1$–$Se_1$ distance is 3.776 Å with a −ICOHP($E_f$) of only 0.024. Together with the geometry, these results identify two strong covalent $Se_1$–$Se_2$ bonds within each V-shaped $Se_1$–$Se_2$–$Se_1$ trimer. $Rb_1$ and $Rb_2$ donate 0.617 and 0.561 electrons, respectively, while $Se_1$ and $Se_2$ gain 0.538 and 0.102 electrons. This charge distribution supports an ionic description involving Rb cations and covalently bonded $Se_3$ polyanions, consistent with the original structural study [20]. The coexistence of covalent $Se_3$ trimers and ionic Rb–Se coordination suggests distinct vibrational responses within the crystal.

Table I. Calculated Bader charge transfer $\Delta q$ in units of the elementary charge, interatomic distances $L_{ij}$ in Å, and −ICOHP($E_f$) values in eV for the inequivalent atomic pairs in $Rb_2Se_3$. Positive $\Delta q$ denotes electron donation. The −ICOHP($E_f$) values are integrated over the occupied states up to the Fermi level.

| **Atoms** | $\Delta q$ | **Bond** | $L_{ij}$ **(Å)** | **-ICOHP($E_f$) (eV)** |
|---|---|---|---|---|
| $Rb_1$ | 0.617 | $Rb_1$-$Se_1$ | 3.441 | 0.353 |
| | | | 3.528 | 0.264 |
| | | | 3.536 | 0.327 |
| | | $Rb_1$-$Se_2$ | 3.407 | 0.280 |
| $Rb_2$ | 0.561 | $Rb_2$-$Se_1$ | 3.414 | 0.349 |
| | | | 3.509 | 0.356 |
| | | | 3.569 | 0.255 |
| | | $Rb_2$-$Se_2$ | 3.380 | 0.241 |
| | | | 3.699 | 0.249 |
| $Se_1$ | -0.538 | $Se_1$-$Se_1$ | 3.776 | 0.024 |
| $Se_2$ | -0.102 | $Se_1$-$Se_2$ | 2.411 | 3.727 |

## B. Phonon dispersion and atom-resolved vibrations

The phonon dispersion and atom-projected density of states provide a basis for distinguishing the vibrations associated with the two bonding environments. Figures 2(a) and 2(b) show the Brillouin zone, the high-symmetry path, and the calculated phonon dispersion. No appreciable imaginary branches appear along this path, supporting harmonic dynamical stability and agreeing with the earlier high-throughput prediction [21]. The $r^2$SCAN calculation retains the main spectral features obtained with PBEsol, including the separation of the four upper optical branches from the lower manifold (Fig. S4 of the Supplemental Material).

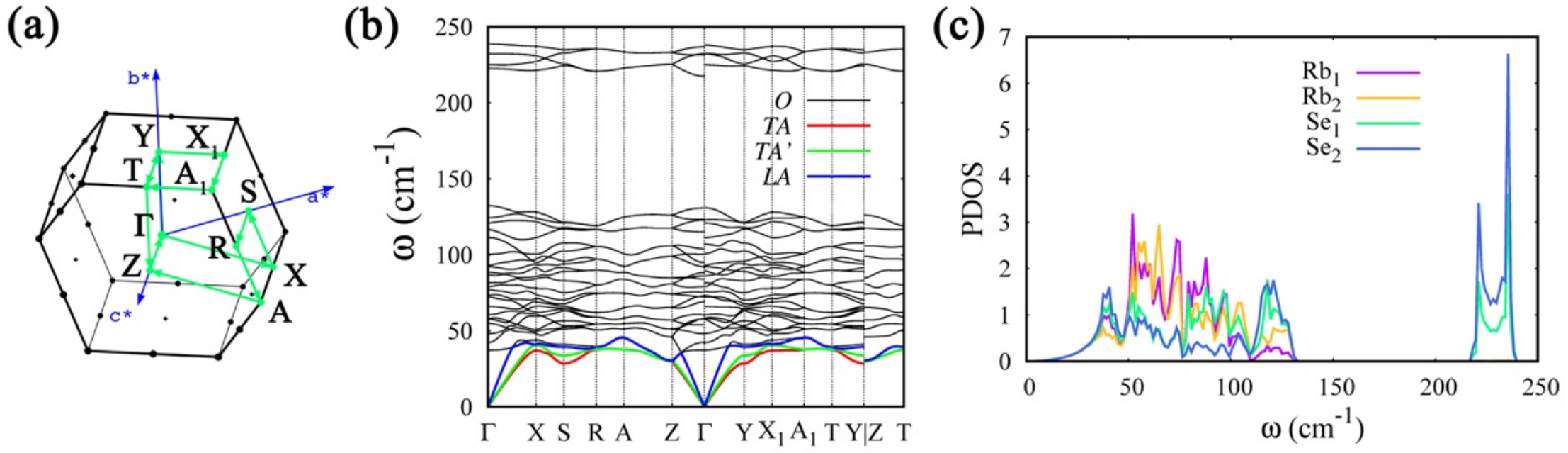

Figure 2. (a) First Brillouin zone and high-symmetry points, (b) phonon dispersion, and (c) atom-projected phonon density of states (PDOS) of $Rb_2Se_3$.

The atom-projected phonon density of states in Fig. 2(c) separates into two frequency regions. A dense manifold below approximately 135 $cm^{-1}$ contains contributions from all four atomic sites. Four optical branches at approximately 215–240 $cm^{-1}$ are dominated by Se motion, with little Rb participation. There is a gap of about 80 $cm^{-1}$ between the two frequency regions. Similar spectral separation has been reported in related compounds containing strongly bonded chalcogen units [57,58].

The harmonic force constants connect this spectral separation to the bonding hierarchy. To compare the local harmonic couplings, we evaluated the Frobenius norm of each $3 \times 3$ interatomic force-constant block, defined as the square root of the sum of its nine squared components. The nearest $Se_1$–$Se_2$ pair has a norm of 7.07 eV/Å$^2$, compared with 0.96 eV/Å$^2$ for the shortest Rb–Se contact (Fig. S5 of the Supplemental Material). This difference is consistent with high-frequency internal vibrations of the covalent trimers and lower-frequency motion involving both Rb and Se in the surrounding coordination framework. At Γ, the four highest-frequency modes have 99.40–99.95% of their mass-weighted eigenvector norm on Se. The longitudinal component accounts for 72.4–83.3% of the squared relative displacements across the intratrimer bonds (Fig. S6 and Table SII of the Supplemental Material), identifying these modes as predominantly Se–Se stretching. Together, the projected density of states, harmonic couplings, and Γ-point eigenvectors support the association of the upper manifold with internal trimer vibrations.

## C. Trimer vibrations and the pseudo-atom-chain arrangement

To complement the COHP analysis with a real-space view of the electronic structure, we calculated the two-dimensional (2D) electron localization function (ELF) [43] in the (100), (010), and (001) plane [Figs. 3(a)-3(c)]. The maps show pronounced electron localization around the Se sites, providing a real-space picture of the trimers identified above. The atomic coordinates show that the trimer planes are approximately parallel to (041) and $(04\bar{1})$, each tilted by about 19.6° from (010). The two intratrimer $Se_1$–$Se_2$ bonds extend close to the [101] and $[10\bar{1}]$ directions, with substantial components along [100] and [001] and smaller components along [010].

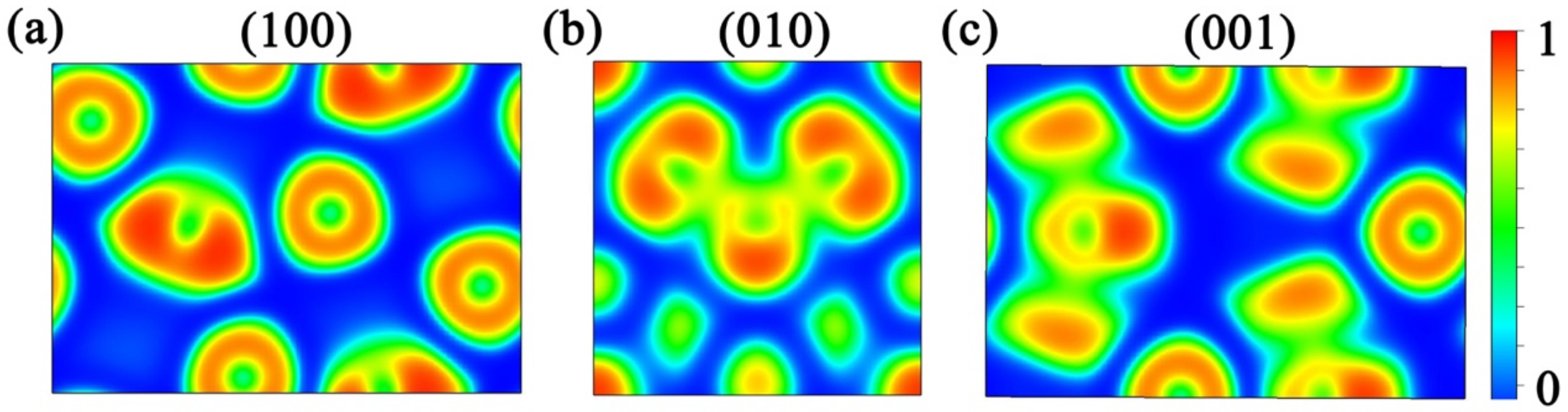

Figure 3. The two-dimensional (2D) electron localization function (ELF) of $Rb_2Se_3$ in the (a) (100), (b) (010), and (c) (001) plane, respectively.

Representing each $Se_1$–$Se_2$–$Se_1$ trimer as a pseudo-atom X reveals how these units are arranged within the crystal (Fig. 4). The resulting structure exhibits a pseudo-one-dimensional chain arrangement along [001] within the (100) plane. This representation highlights the directional organization of the trimers within the three-dimensional ionic coordination network. A related coexistence of strongly bonded structural units and softer surrounding motion has been examined in quasi-one-dimensional $A_2PdPS_4I$ (A = K, Rb, Cs), where covalent chains and cooperative vibrations of the interchain atoms contribute differently to the lattice dynamics [59]. The arrangement in $Rb_2Se_3$ similarly motivates examining how the internal bonds and surrounding coordination contribute to the directional displacement response.

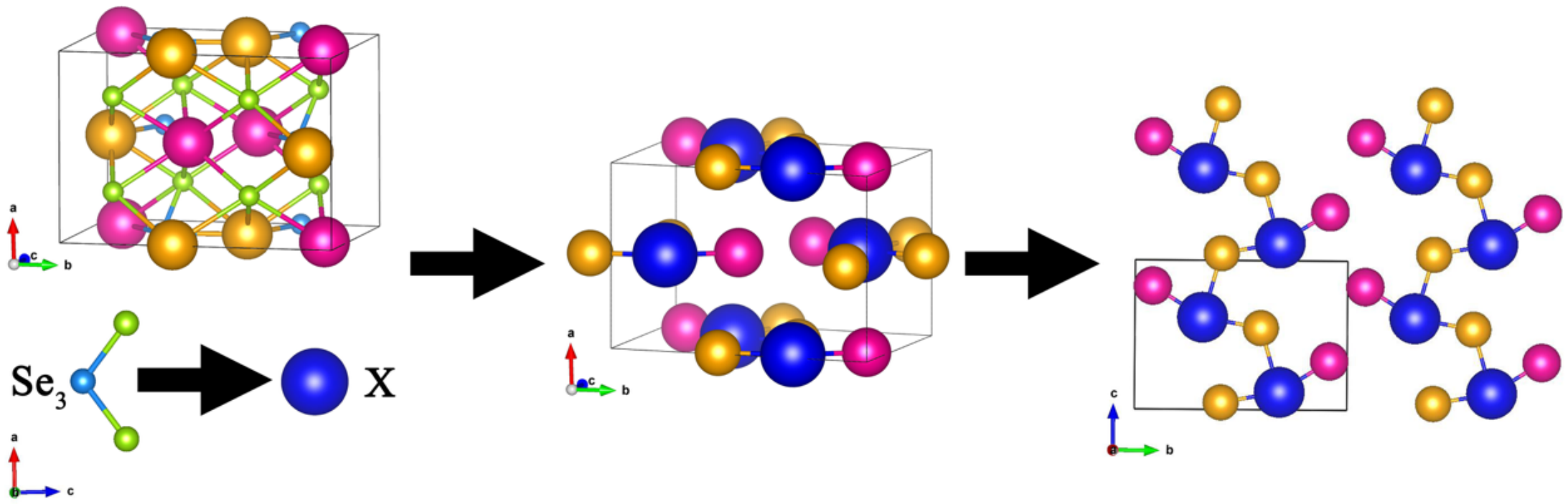

Figure 4. Simplified structure of $Rb_2Se_3$ obtained by representing each $Se_1$–$Se_2$–$Se_1$ trimer as a pseudo-atom (X).

The potential-energy curves in Fig. 5(a) resolve this response for displacements of the four inequivalent atoms along the crystallographic axes. Both Rb sites have their shallowest potentials along [100], perpendicular to the (100) plane containing the pseudo-atom chains. Both Se sites instead have their shallowest potentials along [010], approximately normal to the trimer planes. The latter observation is consistent with the smaller projections of the intratrimer bonds along [010], which reduce the first-order bond-length changes induced by displacements in this direction. The different soft directions of Rb and Se show that the local response depends on the atomic site and its bonding environment. The shallow potentials on both sublattices are consistent with their contributions to the low-frequency phonon manifold.

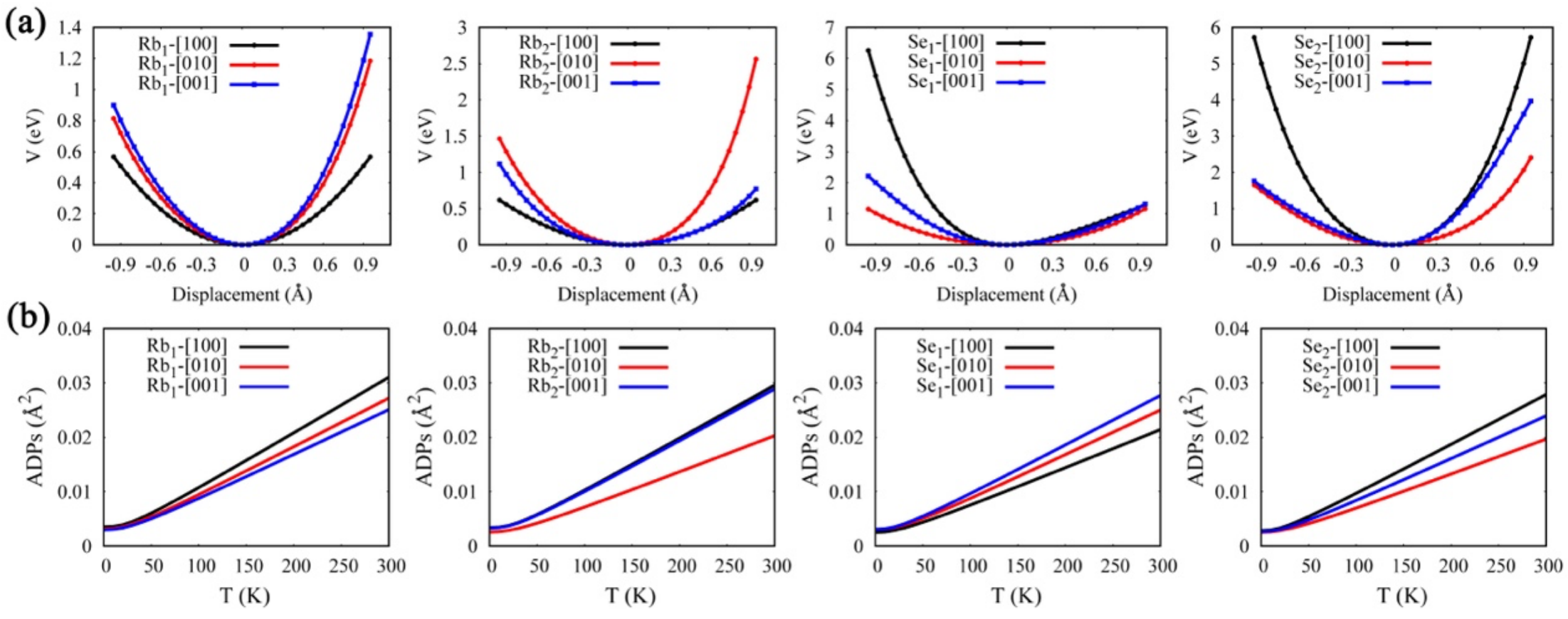

Figure 5. Calculated (a) potential-energy changes V for displacements from equilibrium and (b)

temperature-dependent atomic displacement parameters (ADPs) along [100], [010], and [001] in $Rb_2Se_3$. The columns correspond to $Rb_1$, $Rb_2$, $Se_1$, and $Se_2$, from left to right.

The atomic displacement parameters (ADPs) complement these local energy profiles by describing the mean-square amplitudes of collective lattice vibrations [60]. At 300 K, the largest directional ADP at each inequivalent site exceeds 0.025 Å² [Fig. 5(b)]. These values are larger than those reported for the rattling-like Cu and Cs sites in o-$CsCu_5S_3$, 0.019 and 0.021 Å², respectively [26]. For Rb, the directions with shallower potentials also exhibit larger ADPs. $Se_2$, however, retains a large ADP along [100] despite its relatively steep single-site potential. The two quantities probe different motions. The potential curves describe constrained single-site displacements, whereas the ADP includes the frequencies and displacement patterns of collective normal modes. In a collective mode, correlated displacements can allow substantial Se motion with smaller changes in the internal bond lengths. The ADPs and potential-energy curves therefore provide complementary information about collective vibrational amplitudes and local directional stiffness.

## D. Mode-resolved Grüneisen parameters

To examine how the internal trimer vibrations and collective lattice modes respond to changes in volume, we calculated the mode-resolved Grüneisen parameters using finite-volume phonon calculations in PHONOPY and a third-order-IFC-based evaluation in ShengBTE. The PHONOPY calculations used structures at the equilibrium volume and at volumes scaled by approximately ±3%. The two distributions are shown in Figs. 6(a) and 6(b), respectively. Positive and negative Grüneisen parameters correspond to phonon softening and hardening upon expansion along the structural path considered.

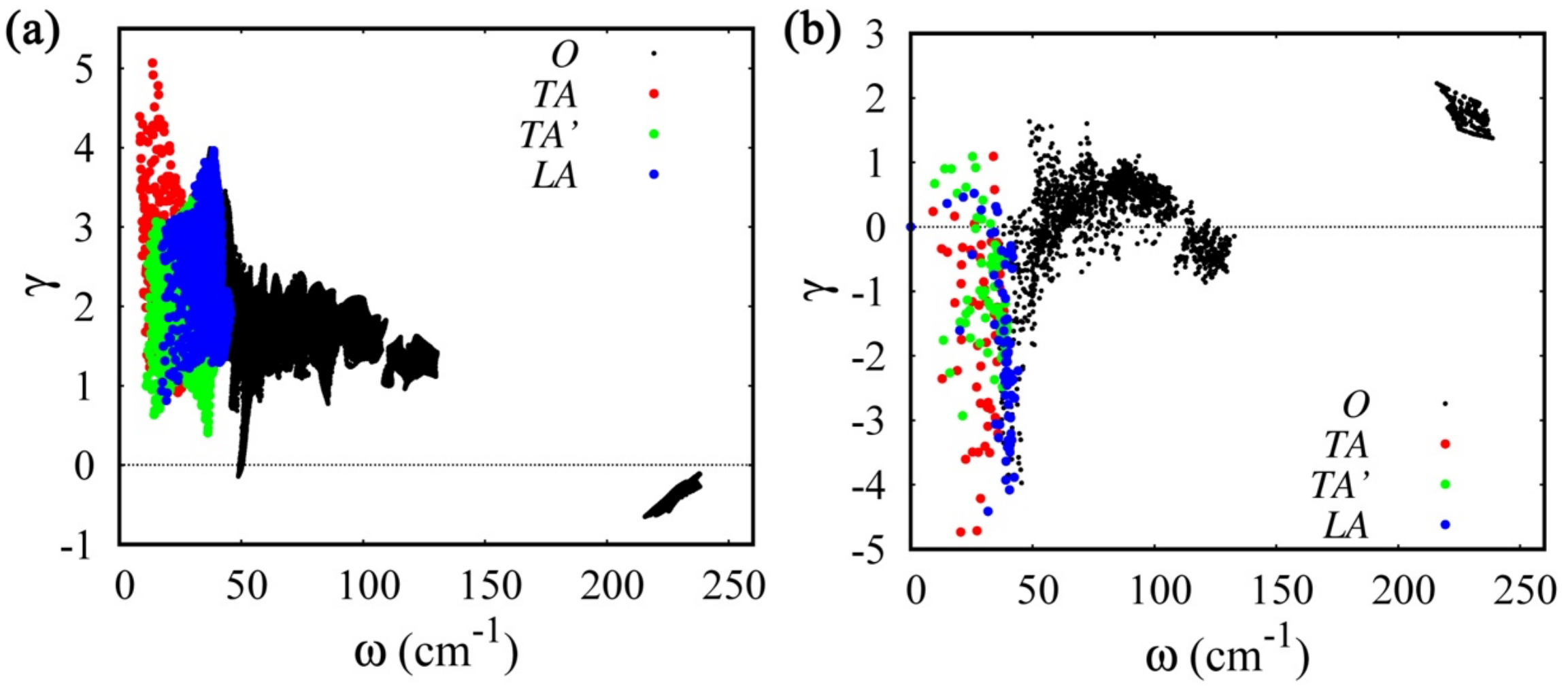

Figure 6. The mode-dependent Grüneisen parameters ($\gamma$) versus phonon frequency in $Rb_2Se_3$, obtained from (a) finite-volume PHONOPY calculations and (b) third-order interatomic force constants.

The finite-volume PHONOPY results show substantial positive Grüneisen parameters for some low-frequency modes, comparable in magnitude to those reported for strongly anharmonic SnSe [7], indicating pronounced volume sensitivity. The lower-frequency manifold is predominantly positive, whereas all four branches in the isolated high-frequency manifold have negative values ranging from −0.651 to −0.109. The third-order-IFC results show a different distribution: positive and negative values coexist in the lower manifold, with modes below 40 $cm^{-1}$ predominantly negative, while the four high-frequency branches are entirely positive, with values between 1.375 and 2.225. Thus, each calculation distinguishes the volume response of the lowest-frequency modes from that of the isolated upper modes, although their predominant signs are reversed between the two calculations.

This frequency dependence connects the volume response to the vibrational character established above. The low-frequency modes involve collective Rb and Se motion within the ionic coordination framework, while the upper modes are associated mainly with internal Se–Se stretching. The largest positive IFC-based parameters occur in the upper manifold, although some low-frequency modes have negative parameters of greater absolute magnitude. Appreciable volume sensitivity therefore occurs in both the framework motion and the internal trimer vibrations. Interpreting the sign difference also requires considering the structural response sampled by each calculation. The finite-volume results follow structures with changes in cell shape and internal coordinates, whereas the IFC-based evaluation uses the force constants around the equilibrium structure.

The volume-dependent geometry helps explain the negative finite-volume parameters of the upper modes. Across the calculated volume range, the $Se_1$–$Se_2$–$Se_1$ angle increases from 102.38° to 103.69°, while the intratrimer $Se_1$–$Se_2$ bonds shorten from 2.4145 to 2.4085 Å. In the same deformed cells, holding the equilibrium fractional coordinates fixed instead produces bond extension [Fig. S7(a)]. The four highest-frequency modes at Γ all harden upon expansion [Fig. S7(b)], consistent with their predominantly stretching character and negative finite-volume Grüneisen parameters. The trimers thus accommodate macroscopic expansion through changes in bond angles and internal coordinates that permit local bond contraction. This response shows how stiff covalent units can remain sensitive to the surrounding lattice geometry and links the anharmonic volume response of $Rb_2Se_3$ to the coexistence of covalent trimers and a soft ionic coordination framework.

## E. Lattice thermal conductivity and coherent heat transport

The bonding and vibrational analyses identify stiff internal trimer vibrations together with soft collective motion involving Rb and Se. To determine how these vibrations affect heat transport, we compare four transport models. Models (i) and (ii) include three-phonon scattering and give the particle-like conductivity, $\kappa_p^{3ph}$, and the total conductivity, $\kappa_l^{3ph} = \kappa_p^{3ph} + \kappa_c^{3ph}$, respectively. Models (iii) and (iv) additionally include four-phonon scattering and give $\kappa_p^{3ph+4ph}$ and $\kappa_l^{3ph+4ph} = \kappa_p^{3ph+4ph} + \kappa_c^{3ph+4ph}$, respectively. Isotope scattering is included in both scattering settings, and each coherent contribution is evaluated using the corresponding phonon linewidths. Figure 7 shows the temperature dependence of the lattice thermal conductivity calculated using the four models along the three crystallographic axes, while Table II summarizes their values at 300 K.

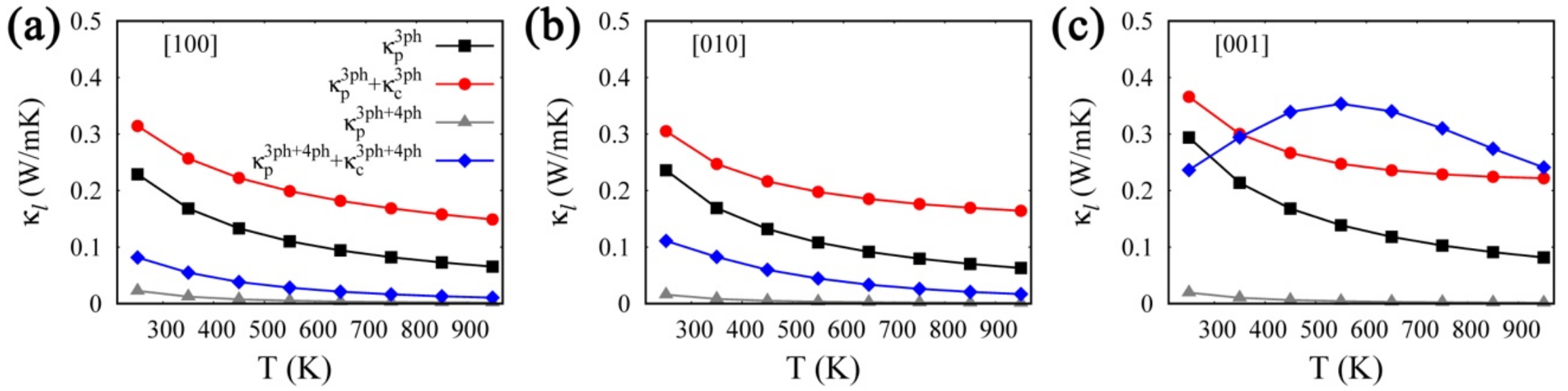

Figure 7. The calculated temperature-dependent lattice thermal conductivity along (a) [100], (b) [010], and (c) [001] directions in $Rb_2Se_3$ for the four transport models defined in the text.

**Table II.** Calculated lattice thermal conductivity contributions at 300 K along [100], [010], and [001]. All conductivities are in W/mK. Models (i) and (ii) include three-phonon scattering; models (iii) and (iv) include three- and four-phonon scattering. In both scattering models, $\kappa_l = \kappa_p + \kappa_c$.

| **Model** | [100] | [010] | [001] |
|---|---|---|---|
| (i) $k_p^{3ph}$ | 0.19382 | 0.19696 | 0.24726 |
| (ii) $k_p^{3ph} + k_c^{3ph}$ | 0.28123 | 0.27091 | 0.32694 |
| (iii) $k_p^{3ph+4ph}$ | 0.01628 | 0.01092 | 0.01359 |
| (iv) $k_p^{3ph+4ph} + k_c^{3ph+4ph}$ | 0.06659 | 0.09634 | 0.26489 |

With three- and four-phonon scattering included, the total conductivities at 300 K are 0.067, 0.096, and 0.265 W/mK along [100], [010], and [001], respectively. The lowest value is below the measured cross-grain conductivities of five hardwood species, approximately 0.12–0.18 W/mK [61], and all three values are well below the room-temperature thermal conductivity of soda–lime–silicate float glass, approximately 1.0 W/mK [62]. Table II shows that these low values combine strongly suppressed particlelike conduction with a substantial coherent contribution.

Four-phonon scattering reduces the particle-like conductivity by 91.6%, 94.5%, and 94.5% along [100], [010], and [001], respectively. The remaining values are only 0.0163, 0.0109, and 0.0136 W/mK, compared with the corresponding three-phonon values of 0.1938, 0.1970, and 0.2473 W/mK. Particlelike conduction is therefore strongly suppressed along every axis, with a residual maximum-to-minimum ratio of only 1.49.

The coherent contribution responds differently. At 300 K, adding four-phonon scattering increases it from 0.0797 to 0.2513 W/mK along [001], produces a smaller increase along [010], and decreases it along [100]. In the Wigner description, scattering also modifies the spectral overlap between phonon branches, allowing coherent transfer to respond differently from particle-like conduction [51] [52]. The coherent contribution consequently accounts for 75.6%, 88.7%, and 94.9% of the total conductivity along the three axes. Its maximum-to-minimum ratio reaches 4.99, showing that coherent transfer determines most of the conductivity anisotropy. Its largest contribution follows [001], the pseudo-atom-chain direction identified in Fig. 4. This balance differs from that in the predicted quasi-one-dimensional $BiI_3$ phase, where suppressed phonon coherence contributes to ultralow conductivity [63]. The comparison highlights the importance of evaluating the coherent channel explicitly in crystals with chainlike structural motifs.

The cumulative conductivities at 300 K locate the frequency ranges associated with this directional response (Fig. S8). With both scattering orders included, the particle-like conductivity accumulates mainly within the lower manifold along all three axes. The upper manifold, however, accounts for approximately 64% of the cumulative coherent conductivity along [001], compared with approximately 10% along [100] and 3% along [010]. These fractions are obtained from the increment above 215 $cm^{-1}$, within the spectral gap, relative to the final cumulative value. They follow the frequency allocation of the coherent output and do not resolve individual coupled branch pairs. The pronounced upper-frequency increment along [001] associates the trimer-dominated spectral region with the direction of the largest coherent heat current.

The temperature dependence further demonstrates the influence of coherent transfer (Fig. 7). The particle-like conductivity decreases throughout the sampled 250–950 K range. The total conductivity also decreases along [100] and [010], but along [001] it rises from 0.236 W/mK at 250 K to a sampled maximum of 0.354 W/mK at 550 K, before decreasing to 0.241 W/mK at 950 K. This nonmonotonic behavior follows the coherent contribution. Moreover, from 450 K onward on the sampled temperature grid, the [001] total conductivity including four-phonon scattering exceeds the three-phonon total. Additional scattering can therefore suppress particle-like conduction while increasing the total conductivity through its effect on coherent transfer. The coherent channel thus governs both the strong anisotropy and the nonmonotonic temperature dependence along the pseudo-atom chains.

## F. Four-phonon scattering mechanisms

The strong reduction of particle-like conductivity raises the question of which vibrations and scattering processes are primarily involved. Figure 8 compares the frequency-dependent three- and four-phonon phase spaces and scattering rates at 300 K. The phase spaces describe processes permitted by energy and momentum conservation, while the rates additionally depend on phonon occupations and anharmonic matrix elements [50]. The Brillouin-zone-weighted four-phonon rate sum is approximately 10.2 times the three-phonon sum, demonstrating the substantial additional broadening introduced by four-phonon processes.

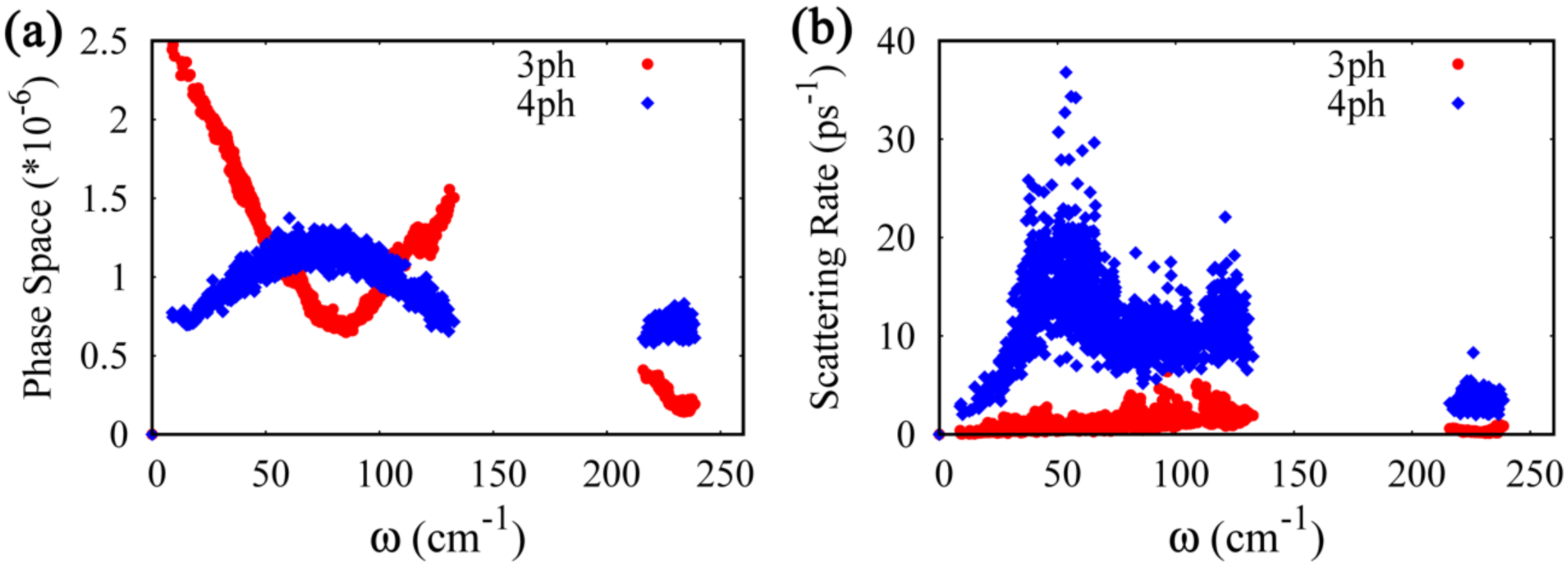

Figure 8. Frequency-dependent (a) unweighted phase spaces and (b) scattering rates for three- and four-phonon processes in $Rb_2Se_3$ at 300 K.

Most of this four-phonon broadening occurs within the lower manifold, where Rb and Se participate in collective motion. Modes between 40 and 130 cm$^{-1}$ account for 86.2% of the total weighted four-phonon rate sum, whereas the four isolated upper branches contribute only 4.3%. The principal broadening therefore affects collective framework vibrations. This distribution complements the cumulative conductivity results discussed above, where the upper manifold contributes substantially to coherent heat transfer along [001]. The frequency range carrying a large coherent contribution is thus distinct from the range accounting for most of the four-phonon rate sum.

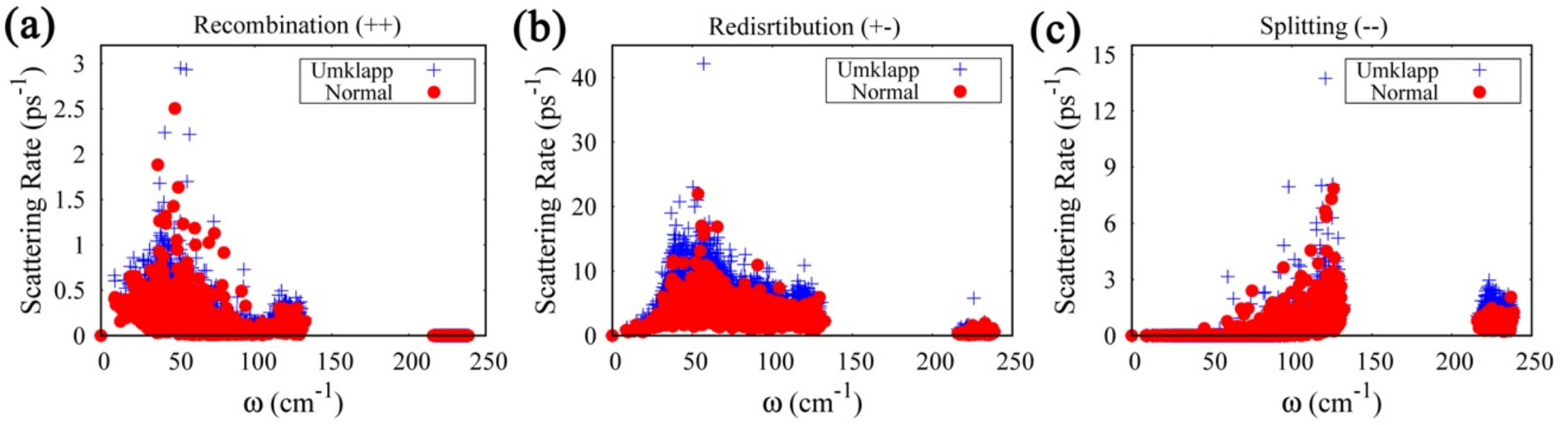

Figure 9. Mode-resolved four-phonon scattering rates of $Rb_2Se_3$ at 300 K, resolved into Normal processes (red circles) and Umklapp processes (blue plus signs): (a) Recombination (++) channel; (b) redistribution (+-) channel; (c) splitting (--) channel.

The channel decomposition in Fig. 9 identifies redistribution (+−) as the dominant four-phonon process, accounting for 87.8% of the weighted rate sum. Recombination (++) and splitting (−−) contribute 3.2% and 8.9%, respectively. Redistribution exchanges energy between two incoming and two outgoing modes and contributes strongly across the populated lower manifold. On the sampled mesh, splitting becomes nonzero above approximately 45 $cm^{-1}$, while recombination is negligible within the isolated upper manifold.

Normal and Umklapp processes both contribute to the calculated broadening. Their Umklapp-to-Normal rate-sum ratios are 1.88, 2.19, and 1.89 for recombination, redistribution, and splitting, respectively, giving an overall ratio of 2.15. These statistics show a substantial Umklapp contribution to the broadening. Redistribution dominates both the Normal and Umklapp four-phonon rate sums. The weighted rate sums characterize the distribution of scattering events, while their consequences for directional heat flow are captured by the conductivity calculations above.

Together, the transport and scattering results distinguish the roles of the two vibrational manifolds. Collective motion in the lower manifold experiences most of the four-phonon broadening, consistent with the strong suppression of particle-like conductivity. The upper manifold contributes a smaller share of the scattering-rate sum but a substantial share of the cumulative coherent conductivity along [001]. Strong internal Se–Se bonds can therefore coexist with ultralow conductivity because the surrounding coordination network supports collective vibrations subject to intense scattering. Coherent transfer carries most of the remaining heat and is strongest along the pseudo-atom-chain direction. The bonding hierarchy, scattering of collective motion, and directional coherent transfer jointly account for the low conductivity and its anisotropy.

## CONCLUSIONS

In this work, the first-principles bonding and phonon calculations identify a clear hierarchy in $Rb_2Se_3$: strongly covalent V-shaped $Se_1$–$Se_2$–$Se_1$ trimers are embedded in a softer ionic coordination network. Representing each trimer as a pseudo-atom reveals chains along [001] within the three-dimensional crystal. The resulting spectrum contains a dense lower manifold and four separated trimer-stretching branches near 215–240 $cm^{-1}$. Four-phonon scattering, dominated by redistribution, strongly reduces the particle-like conductivity. Within the present Wigner treatment, the total conductivities at 300 K are 0.067, 0.096, and 0.265 W $m^{-1}$ $K^{-1}$ along [100], [010], and [001], with coherent contributions of 75.6%, 88.7%, and 94.9%. The weak direction dependence of $\kappa_p$ contrasts with the nearly fivefold variation in $\kappa_c$, whose maximum lies along the pseudo-atom-chain direction [001]. The ultralow conductivity therefore reflects both strong suppression of particle-like transport and the direction dependence of coherent heat transfer. $Rb_2Se_3$ thus illustrates how stiff internal bonds, strongly scattered collective vibrations, and directional coherent transfer can coexist in a low-conductivity ionic crystal.

## ACKNOWLEDGMENTS

This work was supported by the National Natural Science Foundation of China (Grant No. 12104035 and No. 12304124) and by the Natural Science Research Start-up Foundation of Recruiting Talents of Nanjing University of Posts and Telecommunications (Grant No. NY223205).